# The Role of Transport Agents in MoS$_2$ Single Crystals

*Andrea Pisoni*[*1], *Jacim Jacimovic*[1], *Osor S. Barišić*[2], *Arnaud Walter*[1], *Bálint Náfrádi*[1],

*Phillipe Bugnon*[3], *Arnaud Magrez*[3], *Helmuth Berger*[3], *Zsolt Revay*[4], *László Forró*[1]

[1] Laboratory of Physics of Complex Matter, EPFL, CH-1015 Lausanne, Switzerland

[2] Institute of Physics, Bijenička c. 46, HR-10000 Zagreb, Croatia

[3] Single Crystal Growth Facility, EPFL, CH-1015 Lausanne, Switzerland

[4] Heinz Maier-Leibniz Zentrum (MLZ), Technische Universität München, Forschungsneutronenquelle Heinz Maier-Leibnitz (FRM II), D-85747 Garching, Germany

## Abstract

We report resistivity, thermoelectric power and thermal conductivity of MoS$_2$ single crystals prepared by chemical vapour transport (CVT) method using I$_2$, Br$_2$ and TeCl$_4$ as transport agents. The material presents low-lying donor and acceptor levels, which dominate the in-plane charge transport. Intercalates into the Van der Waals gap strongly influence the inter-plane resistivity. Thermoelectric power displays the characteristics of strong electron-phonon interaction. Detailed theoretical model of thermal conductivity reveals the presence of high number of defects in the MoS$_2$ structure. We show that these defects are inherent to CVT growth method, coming mostly from the transport agent molecules inclusion as identified by Total Reflection X-ray Fluorescence analysis (TXRF) and in-beam activation analysis (IBAA).

Molybdenum disulfide, $MoS_2$, has been known for a long time as a mineral but the first detailed studies were performed in the seventies, during the vivid interest for transition metal dichalcogenides[1]. This class of materials possesses a variety of electronic ground states, such as superconductivity, periodic lattice deformation (charge density waves), Mott transition, exciton formation etc[2-4]. Within this cavalcade of fascinating properties, $MoS_2$ was left aside, since it had turned out to be a simple, indirect band gap ($\Delta$) semiconductor[5] of $\Delta = 1.2$ eV.

The revival of interest in this material has come with the isolation of graphene from graphite. Mechanical exfoliation of $MoS_2$ is also possible, and it still remains the preferred technique for obtaining single or few-layers of this material[6-7]. $MoS_2$ single layers achieved by chemical or mechanical exfoliation have been extensively studied for both scientific and applicative purposes[8-11]. Chemical vapour transport (CVT) is the most used method to grow large area single crystals of transition metal dichalcogenides[12-13]. In order to obtain high quality $MoS_2$ few layers by exfoliation of single crystals, the good quality of the starting CVT-grown material must be ensured. Although other authors have already reported a dependence of the doping sign of $WSe_2$, $MoSe_2$ and $WS_2$ at room temperature on the transport agent used during the CVT growth[12-13] a deeper and systematic study on how transport agents influence the electrical and thermal properties of $MoS_2$ single crystals is still missing. Moreover, a recent work has shown that chlorine doping can drastically reduce the width of the Schottky barrier between few layers of $MoS_2$ and metallic contacts allowing high performances in few layers $MoS_2$ field-effect transistors[9]. This observation demonstrates that inclusion of external atoms in $MoS_2$ can also be advantageous for applications.

We measured resistivity of single crystals grown by CVT method using $I_2$, $TeCl_4$ and $Br_2$ as transport agents. Resistivity was measured parallel to the $MoS_2$ planes ($\rho_{ab}$) and out of planes ($\rho_c$). The possibility to measure $\rho_{ab}$ down to 4 K is already strong evidence that charge transport is not dominated by the charge transfer gap. In fact, in the case of a pristine sample with $\Delta=1.2$ eV the resistivity would be so high that no measurable current would be able to pass through the sample even at room temperature. Both, the in-plane and the out-of-plane $\rho$, follow a $\rho_0 \exp(E_b/k_B T)$ temperature dependence ($k_B$ is the Boltzmann constant, $E_b$ is the impurity level activation energy and $\rho_0$ is a temperature independent constant). Figure 1 presents the Arrhenius plot of $\rho_{ab}$ for a sample grown using $I_2$ as transport agent. The comparison between resistivities measured for different crystals obtained by the same CVT grown with iodine transport agent results in a scattering of $\rho_0$ and $E_b$ values as shown in the inset of Fig. 1. $E_b$ is found in the 20-90 meV range. Due to the high level of doping, in some cases already at 100 K a conduction channel opens in the impurity band, and one observes very weak activation energy (4-5 meV). The low temperature flat part is presumably due to hopping conduction within the impurity band[14].

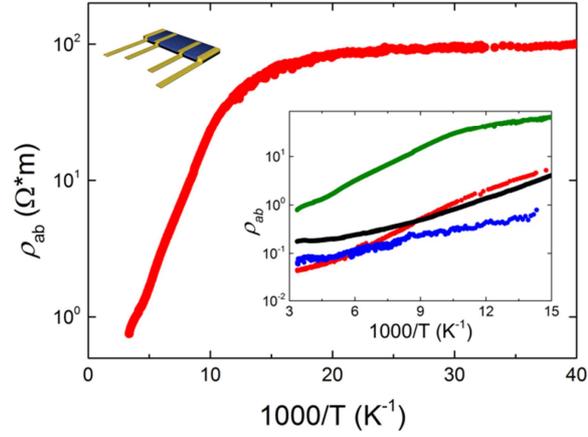

**Figure 1.** Arrhenius plot of the temperature dependence of the in-plane resistivity of a $MoS_2$ single crystal grown using $I_2$ as transport agent. The high temperature slope corresponds to a donor level at 50 meV below the conduction band. The different colour lines in the inset correspond to resistivities measured for four different crystals obtained by the same CVT growth using $I_2$ transport agent.

Figure 2 shows a comparison between the measured $\rho_c$ and $\rho_{ab}$ for the same $MoS_2$ single crystal presented in figure 1. The ratio $\rho_c/\rho_{ab}=100$ at 300 K further increases on cooling. The extracted $E_b = 0.3$ eV for $\rho_c$ is ~6 times higher than for $\rho_{ab}$. In the high temperature regime, the temperature dependent resistivity ratio suggests that the mechanism of conduction is different along the two directions. The higher activation energy for the inter-plane charge transport can be ascribed to intercalated atoms between the $MoS_2$ layers. In fact, $\rho_c$ can be well described by $\rho_0 \exp(E_{int} + E_b/k_b T)$, where the additional energy barrier $E_{int}$ is due to hopping process between nearest neighbours[14]. The charge carriers need phonon assisted hopping to get to the neighbouring layer and this phenomenon manifests itself as an additional activation term in resistivity[14]. Since the phonon assistance is diminishing with the lowering temperature, one observes a strongly increasing resistivity anisotropy.

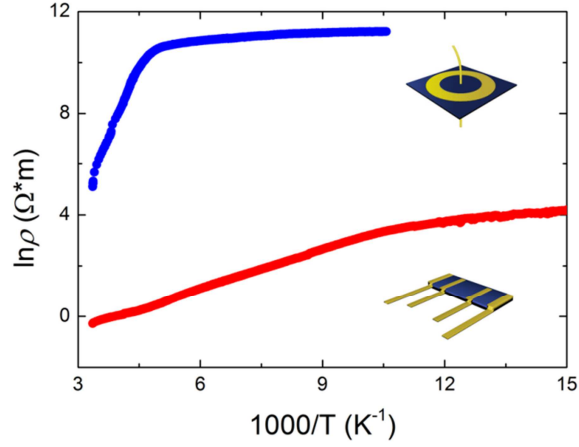

**Figure 2.** Arrhenius plot of the temperature dependence of the in-plane (red curve) and out-of-plane (blue curve) resistivities for a MoS$_2$ single crystal grown using I$_2$ as transport agent.

Whether one deals with a donor or an acceptor impurity level may be deduced from the Seebeck coefficient (*S*). Figure 3 presents the Seebeck coefficient measured parallel to the MoS$_2$ planes for a sample prepared using I$_2$ as transport agent. The sign of *S* suggests that at high temperature the dominant type of charge carriers are hole-like (coming from an acceptor level), while at low temperatures the electron like charge carriers dominate (donor level). Furthermore, at room temperature *S* has a large value of 400 μV/K, typical for non-degenerated semiconductors[15]. In the latter case, *S* is given by

$$S = \frac{k_B}{e}\left(\frac{E_b}{k_B T} + const\right) \quad , \quad (1)$$

where *e* is the electronic charge. This expression gives an increase of |*S*| with decreasing temperature. However, we find that *S* is temperature independent down to 180 K. Such behaviour is characteristic of polaronic charge carriers[16], which have been observed in other

materials like boron carbide[17] and TiO$_2$ anatase[18]. In a simplified picture, the charge carrier excited to the conduction band polarizes the lattice and propagates along the lattice surrounded by this polarization. The energy involved in the creation of this new quasi-particle results with large values of *S*.

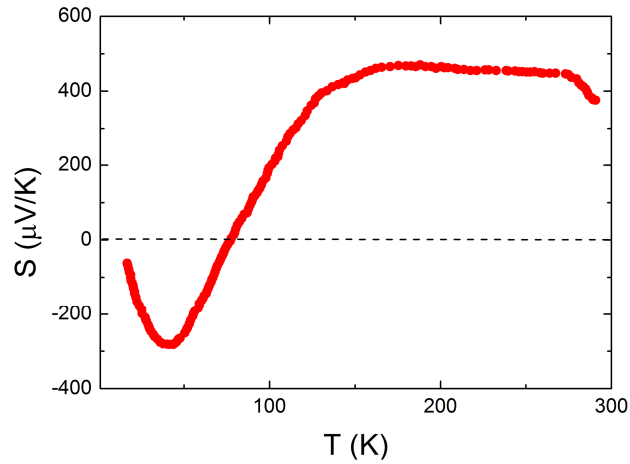

**Figure 3.** Temperature dependence of thermoelectric power measured along the MoS$_2$ planes for the sample grown by I$_2$ transport agent. Electron-like transport at low temperatures and hole-dominated one at high temperatures can be identified.

Thermal conductivity was measured for the MoS$_2$ sample prepared by I$_2$ transport agent. The results are presented in Figure 4. As for other poorly conducting materials, one may expect that the dominant contribution to the thermal conductivity ($\kappa$) is of phononic origin. Indeed, assuming an electron density that justifies the usage of the Wiedemann–Franz law, one obtains a very low estimation for the value of the thermal conductivity due to charge carriers, being much less than 1 W/(Km). Therefore, one can safely neglect such term and analyse $\kappa$ behaviour in terms of purely phonon contribution. At room temperature we measured $\kappa$=15 W/Km, a value that is much lower than in other layered materials like graphite (κ=2000 W/Km)[19]. Room temperature thermal conductivity in a few and single layer MoS$_2$ was measured by a few authors. The measured data span between κ= 30 W/Km and 55 W/Km, depending on the layer thickness and the way in which the samples were grown[20-22]. This is a clear indication that impurities and/or strong anharmonicities play a significant role. In particular, from 10 to 300 K, κ reveals three different temperature intervals: i) the low temperature regime, below 40 K, where κ quickly diminishes as temperature decreases, ii) the intermediate temperature regime, from 40 to 70 K, in which κ exhibits a maximum, and iii) the high temperature part, from 70 to 300 K, in which κ becomes suppressed by temperature. This behaviour has strong resemblances to other semiconducting materials and may be investigated in more details in terms of three independent mechanisms (relaxation times) for phonon scattering: Umklapp phonon scattering, impurity scattering and sample boundary scattering[23]. With these three scattering contributions, in the context of the Callaway formalism, the integral expression for thermal conductivity due to the phonons is given by[23]

$$\kappa = CT^3 \int_0^{\theta_D/T} \frac{1}{a_1 + a_2 T^4 x^4 + a_3 T^3 x^2 \exp[-w_u/T]} \left[ \frac{x^4 e^x}{(e^x - 1)^2} \right] dx \qquad (2)$$

with $C = \left(\frac{k_B}{2\pi^2 v_s}\right)\left(\frac{k_B}{\hbar}\right)^3$, $x = \hbar\omega/k_B T$, $\theta_D$ is Debye temperature, $k_b$ Boltzmann constant, $\hbar$ Planck constant, and $v_s$ the average speed of sound. $a_1$, $a_2$, $a_3$ and $w_u$ are parameters that should be determined from experiments, by using a fitting procedure.

All of the scattering mechanisms in equation (2) are active in the entire temperature range shown in Figure 4, however, not with equal importance. Phonon-phonon interaction is dominant at high temperatures, where, thanks to the thermal excitations, the phonon thermal occupation is high for almost all wavelengths and accompanied with strong anharmonicities. By lowering T, the phonon thermal occupation decreases and Umklapp scattering becomes less significant, which results in an overall rise of thermal conductivity. Yet, by further lowering the temperature, other scattering processes become more efficient. That is, effects of impurities start to dominate at about 1/5 of Debye temperature ($\theta_D = 500$ K for $MoS_2$[24]). At very low temperatures, only the long-wave phonons remain in the system, with a mean free path being constrained by sample dimensions.

The values found for $a_1$, $a_2$, $a_3$ and $w_u$ of equation (2) are given in Table 1. The corresponding fitting curve is shown in blue in Fig. 4, nicely reproducing the experimental data in the entire temperature range. While the Umklapp scattering seems less efficient in comparison to results obtained for a semiconductor with similar values of Debye frequency and κ at room temperature (rutile $TiO_2$)[25], the scattering due to impurities is significantly enhanced for the $MoS_2$ sample. For the latter, $a_2$ is two orders of magnitude higher than for rutile $TiO_2$. This effect is additionally depicted by a low value of κ at the maximum with respect to the room temperature value, i.e., only a factor of 4, compared to a factor 100 observed in pure rutile $TiO_2$[25]. The expected low-temperature $T^3$ behaviour is suppressed up to very low temperature, well below 15 K, which is the lowest temperature investigated in our experiment in Figure 4. All these findings show that our $MoS_2$ sample involves very strong scattering due to impurities. In particular, we believe that this effect should be ascribed to iodine contamination.

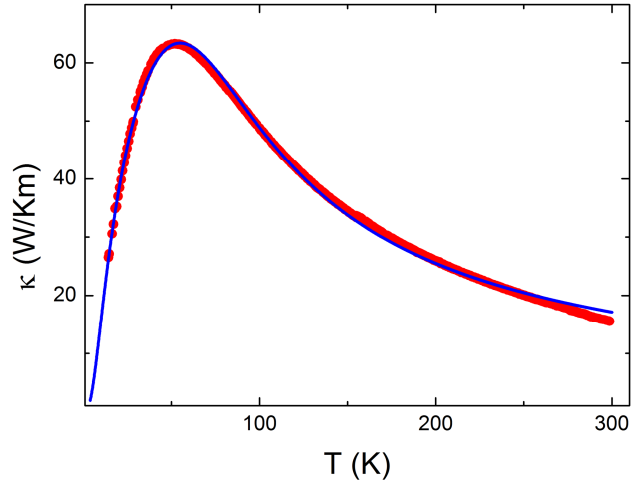

**Figure 4.** Temperature dependence of basal thermal conductivity of $MoS_2$ single crystal obtained using $I_2$ as transport agent. The blue curve represents the fit obtained by using equation (2).

**Table 1.** Fitting parameters by using Callaway's approach

| $a_1/C$ | $a_2/C$ | $a_3/C$ | $w_u/C$ |
|---|---|---|---|
| $1.8 \times 10^6$ | 0.69 | $2.09 \times 10^2$ | 200 |

The results presented so far indicate that the iodine transport agent causes an unintentional doping of the CVT grown $MoS_2$ and that the charge carriers have a polaronic character. The presence of transport agents into the bulk of $MoS_2$ single crystals was confirmed by IBAA and TRXF analysis (see Supporting Information). In the particular case of samples grown using $I_2$, TXRF reveals iodine contents up to 0.48 molar percent. In order to investigate the effect of different transport agents on the electrical properties of $MoS_2$ we studied single crystals produced by $TeCl_4$ and $Br_2$ transport agents. Figure 5 shows the comparison between Arrhenius plots of the in-plane resistivity of $MoS_2$ single crystals grown by the different transport agents. Similarly to the iodine case we measured resistivity of different samples obtained by the same CVT growth using $TeCl_4$ and $Br_2$ as transport agents. The results (not shown here) present the same scattering in $\rho_0$ and $E_b$ as displayed in the inset of Fig. 1. In all the cases resistivity is measurable down to low temperatures and the temperature dependence is very similar, only the absolute value of $\rho$ appears to be transport agent dependent. This could be due to different ways in which transport agents are incorporated into the crystal structure. In fact the layered structure of $MoS_2$ offers interstitial sites[26] and transport agents can intercalates between the $MoS_2$ layers. Iodine incorporation between planes of metal dichalcogenides has already been observed by scanning tunnelling spectroscopy in other compounds grown by CVT method in our laboratories[27]. Iodine can enter into the lattice by substituting S (forming $MoS_{2-x}I_x$) introducing one electron per formula unit. However iodine can also be intercalated in between the $MoS_2$ layers. In this case the chemical formula is $MoS_2I_x$ and iodine induces one hole per formula unit. Although iodine can easily be desorbed from the surface by extensive washing, the removal of substituted or intercalated $I_2$ is more demanding. Probably, high temperature annealing in vacuum or in $H_2S$ vapours would reduce its doping effect.

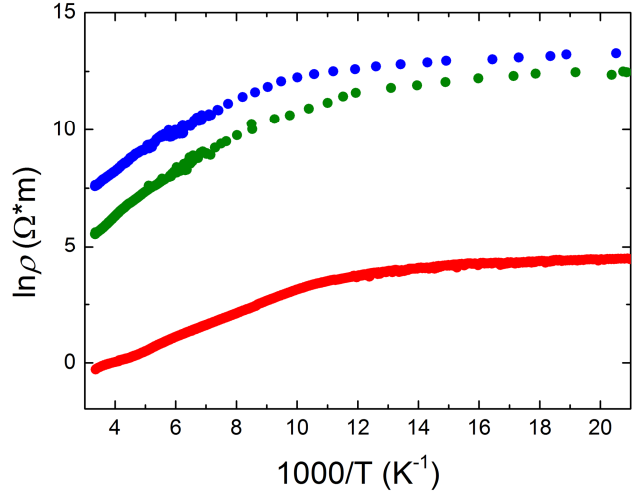

**Figure 5.** Arrhenius plot of the temperature dependence of the in-plane resistivity of $MoS_2$ single crystals grown by $I_2$ (red), $TeCl_4$ (green) and $Br_2$ (blue) as transport agents.

Contrary to the positive sign of Seebeck coefficient of $MoS_2$ grown with $I_2$, single crystals grown with $TeCl_4$ and $Br_2$ transport agents show values of $S$(300K)= -460 µV/K and $S$(300K)= -293 µV/K respectively. Other authors previously reported that CVT growth employing $TeCl_4$ and $Br_2$ transport agents cause n-doping also in $WSe_2$ and $MoSe_2$ crystals[12]. However $WSe_2$ and $MoSe_2$ samples grown with $I_2$ transport agent display p and n-doping at room temperature respectively[12]. Positive Seebeck coefficients were also observed in naturally grown $MoS_2$ crystal[28]. However, none of these samples displayed a change in $S$ at low temperatures. The smooth transition from holes dominated transport to electrons dominated transport in our $MoS_2$ sample grown by $I_2$ indicate that both, p and n type of charge carriers are present in the material. We showed before that iodine intercalation could give extra holes. Sulphur vacancies could be instead the cause of extra electrons, as it was observed in $WS_2$[13].

In conclusion, we have shown that MoS$_2$ single crystals grown by the chemical vapour transport method suffer from an unintentional doping by the transport agent molecules. Temperature dependent electrical and thermal transport measurements clearly reveal the presence of defect and inclusions inside the material, as confirmed by our IBAA and TXRF studies. This shortcoming of CVT grown MoS$_2$ should be always taken into account in the discussion of its physical properties, especially when exfoliated MoS$_2$ single layers are foreseen for applications.

**Experimental Methods**

First, MoS$_2$ powder was synthesized by heating a mixture containing stoichiometric amounts of molybdenum (99.9% pure, Alfa Aesar) and sulphur (99.999% pure, Alfa Aesar) at 1000°C for 7 days in an evacuated and sealed quartz ampoule. The mixture was slowly heated from room temperature to 1000°C for 12 hours, in order to avoid any explosion due to the strong exothermic reaction and the high volatility of sulphur. From this powder, MoS$_2$ crystals were grown using chemical vapour transport (CVT) with iodine, bromine and tellurium tetrachloride as transport agent at ca. 5mg/cm$^3$. All quartz tubes used for vapour transport typically have an inner diameter of 16 mm and a length of 20 cm. The total powder charge is 5 g. A very slight excess of sulphur is always included (typically 0.5 wt% of the charge) to ensure the stoichiometry in the resulting crystals. The excess of sulphur is not incorporated into the dichalcogenide crystals, but condenses as elemental sulphur onto the wall of the quartz tube at the end of the CVT process. The source and growth zones were kept at 1060°C and 1010°C, respectively for 7 days in evacuated and sealed quartz ampoules. After this time the furnace is turned off, a small fraction of the charge is transported towards the colder end of the tube, forming crystals with diameters of about 2—8 mm and thick tens of microns, with exception of $TeCl_4$, where even 400 μm thicknesses where observed. The resulting crystals were washed with acetone and dried in vacuum. X-ray diffraction study has shown that $MoS_2$ obtained in this way belongs to the 2H polymorphism (see supporting information).

Resistivity measurements in the *ab* plane and along the *c*-axis were performed in a conventional four probe configuration after evaporation of chromium/gold contacts. Sketches are shown as insets in Figure 1 and Figure 2. For the $\rho_c$ configuration the outer large circle is the current lead and the point-like contact in the middle is the voltage electrode. For the thermoelectric power measurement the sample was anchored to a ceramic bar on which a temperature gradient (ΔT) was generated by a small heater at one end. The ΔT across the sample was measured by a differential Chromel-Constantan thermocouple attached to the sample (the details are described elsewhere[29]). Thermal conductivity (κ) was measured by the steady state method, using a reference sample to measure the heat current through the $MoS_2$ single crystal as described elsewhere [30].

**Notes**

A. Pisoni and J. Jacimovic contributed equally to this work. The authors declare no competing financial interests.

ACKNOWLEDGMENT

The research was supported by the Swiss National Science Foundation. The technical assistance of Dr. R. Gaál is gratefully acknowledged. We thank to Stefano Pisoni for help in performing resistivity measurements. We thank to N.M. Norbert, M. Cantoni, for help in crystals characterization. This research project has been supported by the European Commission under the 7th Framework Programme through the "Research Infrastructures" action of the Capacities Programme, NMI3-II, Grant Agreement number 283883.